\documentclass[12pt]{article}

\usepackage{scicite}
\usepackage{siunitx}
\usepackage{times}
\usepackage{graphicx}

\newenvironment{sciabstract}{%
\begin{quote} \bf}
{\end{quote}}

\title{Polaritons in photonic hypercrystals of van der Waals materials} 

 \author
{Nihar Ranjan Sahoo,$^{1}$ Brijesh Kumar,$^{1}$ S.S. Jatin Prasath,$^{1}$  Aneesh Bapat,$^{1}$\\ Parul Sharma,$^{1}$ Anshuman Kumar$^{1\ast}$\\
\\
\normalsize{$^{1}$Laboratory of Optics of Quantum Materials,Department of Physics, IIT Bombay, India}\\
\\
\normalsize{$^\ast$E-mail:  anshuman.kumar@iitb.ac.in}
}

\date{}

\begin{document} 


\baselineskip24pt


\maketitle


\begin{sciabstract}
In-plane Hyperbolic Phonon polaritons (HPhPs) are quasiparticles formed via coupling of photons and optical phonons in in-plane hyperbolic materials and offer unique applications in sensing, thermal emitters and high resolution imaging. However, the large momentum mismatch between photons and these in-plane HPhPs has restricted their technological potential as most experimental demonstration rely on sophisticated and expensive near field detection schemes. In this work, using the example of $\alpha-$MoO$_3$, we demonstrate that by constructing photonic hypercrystals of this material, one can not only excite these in-plane HPhPs in the far field but also tune the far field response via twisting the hypercrystal lattice with respect the lattice of $\alpha-$MoO$_3$. Our findings will pave the way for the development of practical in-plane HPhP devices as well as provide access to new fundamental physics of such materials via conventional and well developed far field measurement techniques. 
\end{sciabstract}

\section*{Introduction}

The spectral region of mid-infrared (mid-IR: 3-30 um) or long wavelength infrared (LWIR: 8-15 um) is a rich playground for a multitude of applications that spans across various fields like, from biomedical industry\cite{Xue2019,Oh2018,lahiri2012medical} and sensing\cite{li2018boron,cheng2015ultrasensitive,Martnez2016} to astronomy\cite{akiyama2019first} and IR imaging,\cite{Tong2020,gade2014thermal} and in facilitating free space communication\cite{Hao2017,dang2020should} and several other applications\cite{Pile2012}. Despite the potential benefits of mid-IR or LWIR spectral regions, the development of IR technologies is fraught with critical challenges. These include the requirement for materials with a robust optical response, the need for miniaturization, and the integration of IR optical devices at a chip scale\cite{Law2013,Fang2019}.
Hyperbolic metamaterials\cite{poddubny2013hyperbolic,guo2020hyperbolic,kildishev2013planar}, which exhibit negative real values of the dielectric permittivity in at least one crystal direction while remaining positive along other directions at the same frequency, in-principle enable strong light confinement at the nanoscale level\cite{poddubny2013hyperbolic}, theoretically offer unprecedented optical density of states in the IR spectral region and potential for integration with chip-scale optical devices, delivering several device applications. However, reliance on artificial metamaterials which typically incorporate plasmonic components can result in significant high losses,\cite{tassin2012comparison} limiting their practical application. To this end, recently discovered natural, highly anisotropic van der Waals (vdW) crystals, such as h-BN\cite{dai2014tunable,caldwell2014sub,caldwell2019photonics}, $\alpha$-MoO\textsubscript{3}\cite{zheng2019mid,ma2018plane,de2021nanoscale}, $\alpha$-V\textsubscript{2}O\textsubscript{5}\cite{taboada2020broad,dixit2021mid}, $\beta$-Ga\textsubscript{2}O\textsubscript{3}\cite{passler2022hyperbolic} and others, exhibit low-loss hyperbolic modes arising from the formation of phonon polaritons (PhPs) that enable strong light-matter interactions and ultrahigh confinement on the order of approximately 1/100 of the free space wavelength\cite{dai2014tunable,low2017polaritons,giles2018ultralow}. \par

Phonon polaritons are a unique class of quasiparticles resulting from the coupling of crystal vibrations (phonons) with light (photons), allowing for the confinement of infrared light to deep sub-wavelength scales, significantly enhancing the interactions between light and matter. In the case of the hyperbolic vdW materials, the PhPs arise from the extremely anisotropic lattice vibrations along different principal axes, and thus making these polaritons hyperbolic (HPhPs). Compared to plasmonic metamaterials, the HPhPs in vdW crystals offer several advantages, including ultra low losses\cite{caldwell2015low,ma2018plane,lee2020image} and high quality factor, making them promising for the development of more efficient and effective photonic devices for various applications\cite{li2018boron,biehs2012hyperbolic,baranov2019nanophotonic,dai2015subdiffractional,li2015hyperbolic,zhang2021interface}.
Recent studies have shown that in the case of biaxial vdW crystals( like, $\alpha$-MoO\textsubscript{3} and $\alpha$-V\textsubscript{2}O\textsubscript{5}), the HPhPs are strongly direction-dependent on its basal plane, where the isofrequency contours exhibiting a hyperbolic geometry. This is in contrast to the in-plane isotropic HPhPs observed in uniaxial vdW materials, such as h-BN, where the in-plane isofrequency curves are circular. The in-plane hyperbolicity in vdW crystal provides extremely high polariton momentum on its surface, opening up new possibilities for manipulating and configuring IR light waves and energy flow in new ways at the nanoscale\cite{MartnSnchez2021}.
The in-plane hyperbolic phonon polaritons are fascinating due to their ability to exhibit unique optical phenomena, including negative refraction,\cite{duan2021planar} topological transitions,\cite{duan2021enabling}, light canalization,\cite{zheng2020phonon} towards planner nanophotonics applications\cite{zheng2022controlling} and twist photonics\cite{duan2020twisted,hu2020topological,chen2020configurable} which can be observed directly in real space using near-field nano-imaging techniques. However, practical applications using in-plane hyperbolic vdW crystals have been restricted to the near-field regime due to the large momentum mismatch for far-field excitation.  Previous studies have suggested that this limitation could be overcome by patterning the vdW crystal with materials like hBN,\cite{caldwell2014sub,alfaro2019deeply} graphene,\cite{brar2014hybrid,ju2011graphene,yan2013damping} semimetals,\cite{wang2020van} topological insulators,\cite{di2013observation} and others. 
In this context, our research explores the potential of far-field excitation of phonon polaritons using nanostructures made from in-plane hyperbolic vdW crystals. A photonic hypercrystal consists of periodic structural modulations in a hyperbolic material\cite{PhysRevX.4.041014}. Our focus is on investigating the use of these structures to overcome the momentum mismatch and enable practical device applications using phonon polaritons. Using the example of $\alpha-$MoO$_3$, we demonstrate that by constructing photonic hypercrystals of this material, one can not only excite these in-plane HPhPs in the far field but also tune the far field response via twisting the hypercrystal lattice with respect the lattice of $\alpha-$MoO$_3$. Our study of the interaction between light and these hypercrystals enables a deeper understanding of the underlying physics and paves the way for the practical realization of phonon polariton-based devices.


\section*{RESULTS}

\subsection*{Far-field optical response of $\alpha$-MoO$\textsubscript{3}$ nano structure}

\begin{figure}[h]
    \includegraphics[width=\linewidth]{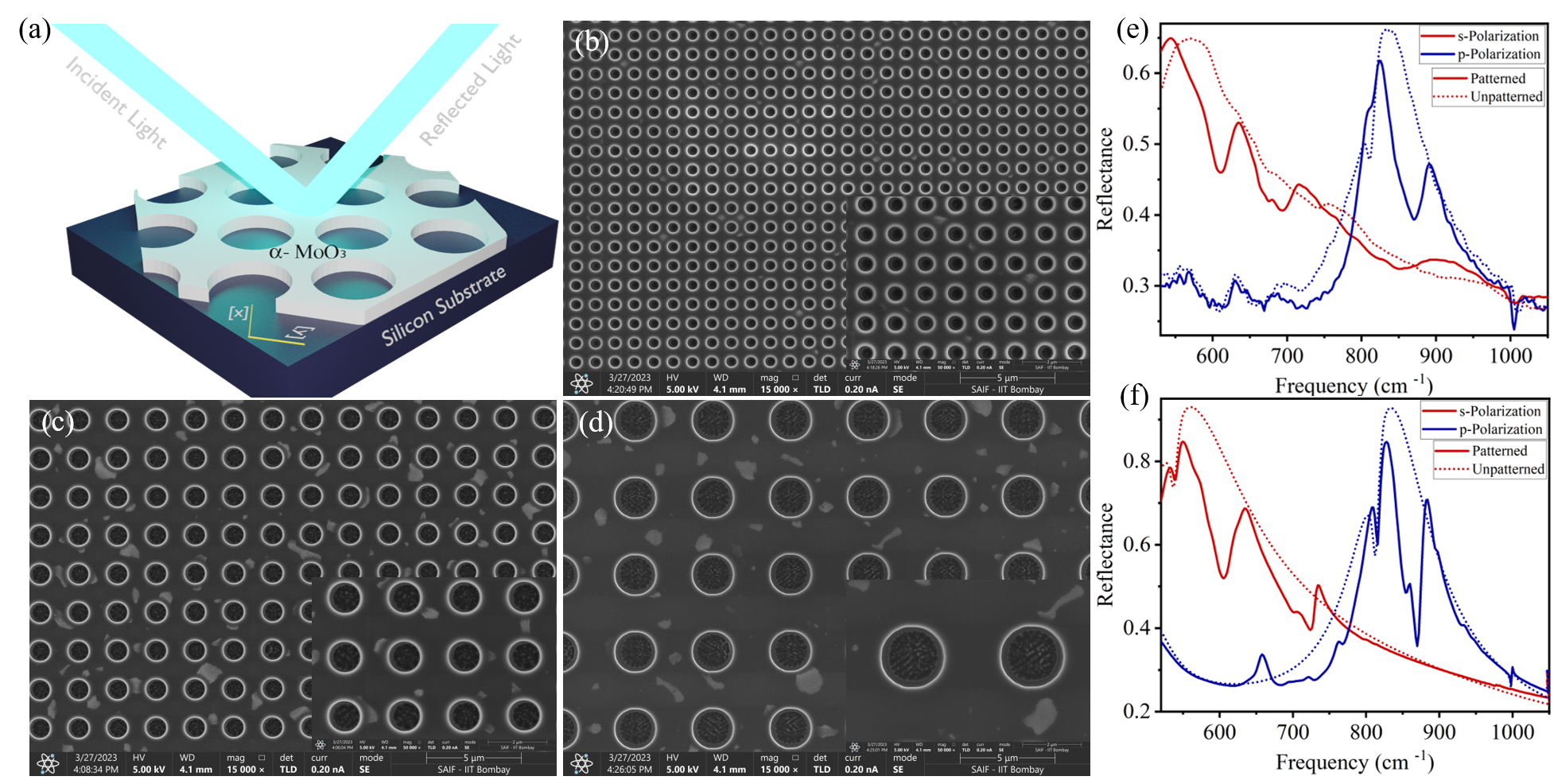}
    \caption{\small \textbf{{Fabrication $\alpha$-MoO$\textsubscript{3}$ hypercrystal}}: (a) Schematic of the far-field excitation of HPhPs due to the $\alpha$-MoO$\textsubscript{3}$ cylindrical nanohole geometry. (b-d), the Scanning Electron Microscopy image of the proposed square-shaped cylindrical nano hole of $\alpha$-MoO$\textsubscript{3}$ with diameters $\SI{0.5}{\micro\metre}, \SI{1}{\micro\metre}$ and $\SI{2}{\micro\metre}$, respectively. The periodicity of the geometry is kept twice the diameter. Inset shows the enlarged SEM image of the proposed hypercrystal. (e-f) shows the experimental and numerically simulated reflectance spectra due to unpatterned (dashed line) and patterned $\alpha$-MoO$\textsubscript{3}$ crystal (solid line). For the comparison patterned sample with a diameter of 1um and periodicity of 2um is considered. The red and blue colour represents incident s- and p-polarized light, respectively     }
    \label{Fig:9}
\end{figure}

In this study, we chose the hypercrystal as a square lattice of cylindrical nano holes in $\alpha$-MoO$\textsubscript{3}$ on a pristine high-resistivity silicon substrate for the far-field excitation of HPhPs (Fig. 1(a)). In our notation, the long crystallographic direction [100] of $\alpha$-MoO$\textsubscript{3}$ was along the $x-$axis of the lab frame and its orthogonal component [001] was along the $y-$axis. The $\alpha$-MoO$\textsubscript{3}$ hypercrystal were fabricated, using Focused Ion Beam (FIB) milling (Method section) where the hole diameter was chosen as $w$, and the periodicity was $2w$. The diameter $w$ was much smaller than the excitation mid-IR wavelength. For this study, we varied the hole diameter as $\SI{0.5}{\um}$, $\SI{1}{\um}$ and $\SI{2}{\um}$ respectively (Fig. 1(b-d)), which were fabricated across a $\SI{200} {\um}$ $\times$ $\SI{200} {\um}$ lateral area on a 590 nm thick $\alpha$-MoO$\textsubscript{3}$ flake. We first explored the far-field optical response of a 590 nm thick unpatterned $\alpha$-MoO$\textsubscript{3}$ (Fig. 1(e) (dashed line)) for a broadband long wavelength IR light (520cm$\textsuperscript{-1}$-1050cm$\textsuperscript{-1}$) using a micro-Fourier Transform Infrared (FTIR) Spectroscopy (Methods section). For s-polarized incident light, we obtained a sharp peak in the reflectance spectra at 545 cm$\textsuperscript{-1}$, and for p-polarized light, the peak was obtained around 821 cm$\textsuperscript{-1}$, respectively. This was due to the IR active TO phonon modes along y[001]- and x[100]- crystal direction of $\alpha$-MoO$\textsubscript{3}$\cite{sahoo2022high}. Additionally, we also obtained a dip in the reflectance spectra at 1004 cm$\textsuperscript{-1}$ for both the incident polarizations, which is attributed due to the LO phonon mode along the z[010]-crystal direction. However, we did not find any spectral evidence for the hyperbolic PhP in the unpatterned $\alpha$-MoO$\textsubscript{3}$, and both the reflectance peaks were due to the TO phonon modes in the respective RS bands. This is due to the large momentum mismatch between the free space photon and polaritons, which prevents the coupling of electromagnetic field to HPhPs.

When the linearly polarized light is incident upon our hypercrystal, the nano hole array will diffract the incident mid-IR light, producing waves with an additional momentum corresponding to its reciprocal lattice momentum. These waves can then hybridize with the phonons to form polaritons, which appear in the spectral signature of the scattered light, enabling the far-field excitation of the HPhPs. As the HPhPs propagate, they form standing wave resonances due to multiple reflections from the edges of the holes. The coupled light fields will either be dissipated by lattice vibrations or radiate back to free space, as evidenced by resonance peaks in the corresponding reflectance spectra. Here the cylindrical nano holes exhibit a distinct behavior compared to other geometry like nano gratings. In this case, standing wave resonance can occur in both directions, depending on the polarization state of the incident light, where as for the case of nano gratting the polariton only propagate perpendicular to the direction of the grating long axis, hence the cylindrical geometry enables far-field excitation of HPhPs in both RS bands providing extra degree of freedom for light manipulate. Fig 1(e) (solid line) shows the reflectance spectra due to a typical cylindrical nano hole with diameter w = $\SI{1} {\um}$, and periodicity  $\SI{2} {\um}$. For the incident s-polarized light, a strong reflectance spectrum (solid red line) has peaks at 545 cm$\textsuperscript{-1}$, 638 cm$\textsuperscript{-1}$ and 720 cm$\textsuperscript{-1}$, respectively. The peak at 545 cm$\textsuperscript{-1}$ originates from the TO phonon mode along the $y$-crystal direction, whereas the other two peaks are due to the excitation of HPhPs in the RS band 1. On the contrary, for the incident p-polarized light, there are two distinct peaks coming at 820 cm$\textsuperscript{-1}$ and at 910 cm$\textsuperscript{-1}$ respectively. The first peak is due to the TO phonon mode along the x-crystal direction and the other one is coming due to the excitation of HPhPs in the RS band 2.

 \par

\begin{figure*}[h!]
    \includegraphics[width=\linewidth]{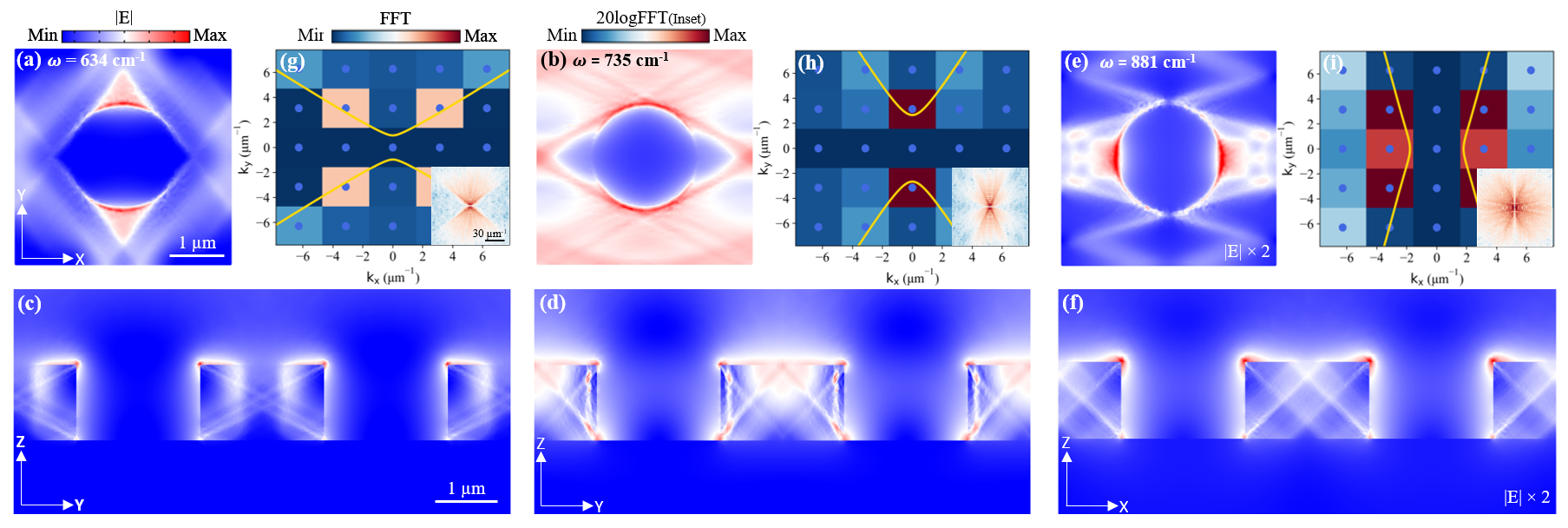}
    \caption{\small \textbf{{Propagation of polaritons in  $\alpha$-MoO$\textsubscript{3}$ hypercrystal}}: 
    (a-b) The normalized electric field profile ($|E|$) in the $xy$ plane for band 1 with $y$-polarized incident light (normally incident) at frequencies 634 cm$\textsuperscript{-1}$ and 735 cm$\textsuperscript{-1}$, respectively. The $xy$ plane is located 5 nm above the $\alpha$-MoO$\textsubscript{3}$ nanostructure. Panels (c-d) illustrate the field profile for the same frequencies but in the $yz$ plane, which passes through the centre of the unit cell. Here, two unit cells are considered to demonstrate the polariton propagation. For band 2, the field profile $|E|$ is plotted at 881 cm$\textsuperscript{-1}$, in (e-f) on the $xy$ and $xz$ planes, respectively, for $x$-polarized incident light. (g-i) shows the nature of analytically calculated IFCs (yellow) for three different frequencies corresponding to Fig. (a-b,e). These hyperbolic IFC are plotted on top of respective FFT image of the corresponding electric field profile ($\text{Re}\{E_z\}$) for all the resonance frequencies in both RS bands, showing intersecting of IFCs with higher intensity FFT giving different mode orders. The blue dots represent the reciprocal space lattice points by considering a periodic lattice with periodicity $\SI{2}{\um}$. Further, the inset represents the FFT of the simulated field profile ($\text{Re}\{E_z\}$) for higher $\textit{k}$ value, where the scale bar is $\SI{30}{\um}$$\textsuperscript{-1}$. For the simulation $\alpha$-MoO$\textsubscript{3}$ unit cell with cylindrical nano hole with a diameter of $\SI{1}{\um}$, a periodicity of $\SI{2}{\um}$, and a thickness of 590 nm is considered.}
\end{figure*}

To gain a deeper understanding of the origin of the reflectance peak observed in our experiments, we conducted numerical simulations using \textsc{comsol multiphysics}. By replicating the experimental setup shown in Fig. 1(a), we obtained similar results in our simulation shown in Fig 1(f), where we plotted the reflectance spectra for a 590 nm thick $\alpha$-MoO$\textsubscript{3}$ cylindrical nano holes (diameter \SI{1}{\micro\metre}) using both $s$($y$)- and $p$($x$)-polarized (solid line) electric field at normal incidence. We also included the reflectance spectra of unpatterned $\alpha$-MoO$\textsubscript{3}$ as a reference, represented by dashed lines. As observed in the experimental results in Fig. 1(e), no additional reflectance peaks were found in the unpatterned geometry, with only the reflectance peak that can be attributed to the TO phonon frequency. The simulation results of the patterned $\alpha$-MoO$\textsubscript{3}$ matched closely with the experimental results, validating the additional resonances observed in the reflectance spectra. However, a small discrepancy was found between the experimental and simulated reflectance due to the irregular structure of the fabricated $\alpha$-MoO$\textsubscript{3}$ cylindrical nano discs compared to the ideal nano discs in the simulation model.\par

To analyze the nature of these modes observed in our simulations, we conducted further analysis by plotting the electric field profile distribution for all the reflectance peaks detected in Fig. 1(f) for both polarization states of incident light. Fig. 2(a-d) shows the normalized electric field ($\mid$E$\mid$) profiles for a $y$-polarized excitation, in both $\textit{xy}$ and $\textit{yz}$ planes for selected resonance frequencies in RS band 1 (where the electric field is along the negative permittivity axis). We observe a 'cross hatched' pattern for the frequencies of 634 cm$\textsuperscript{-1}$ and 735 cm$\textsuperscript{-1}$, indicating the directionality of the HPhPs propagation. Moreover, the angle between the high field crossing line and $z$-direction is consistent with the theoretically calculated angle, $ \theta(\omega)=\frac{\pi}{2}-\arctan\left(\sqrt{\frac{\epsilon_{z}(\omega)}{{-\epsilon_{y(x)}(\omega)}  }} \right) $ which supports the idea that these PhPs originate due to the extreme anisotropic hyperbolic  behavior of the $\alpha$-MoO$\textsubscript{3}$ sample\cite{caldwell2014sub}. In contrast, we do not observe any such zig-zag pattern at the TO phonon frequency (545 cm$\textsuperscript{-1}$), confirming the absence of a hyperbolic polariton resonance (See Supplementary information). Similar plots were obtained for $x$-polarized incident light in RS band 2 (Fig. 2(e-f)) where the electric field is along the negative permittivity direction. This confirms that these hyperbolic polaritons can be excited in the far field for both $x$- and $y$-polarized incident light in the respective RS bands. Our simulation results are consistent with the experimental results obtained in Fig. 1(e), which suggests that the additional resonances observed in the reflectance spectra are due to HPhPs.

Furthermore, the unique in-plane biaxial hyperbolic properties intrinsic to $\alpha$-MoO$\textsubscript{3}$ within its surface plane render it an ideal candidate for exploring polaritonic waves. These waves exhibit a hyperbolic dispersion, classifying them as hyperbolic polaritons. The evidence for this classification becomes evident when we analyze the Isofrequency Contours (IFC) at different polaritonic frequencies in Fig. 2(g-i). These IFCs (Methods section) depict an in-plane open hyperbola shape, with the major axis oriented along ${k_y}$ for frequencies 634 cm$\textsuperscript{-1}$ and 735 cm$\textsuperscript{-1}$ (RS band 1), and along ${k_x}$ for 881 cm$\textsuperscript{-1}$, as HPhPs inhabit RS band 2, respectively. These IFCs of HPhPs, are superimposed atop the Fourier transform (FFT) image of the simulated electric field profile (real part of $E_z$). The FFT is strategically expanded to reveal different modes, evident in specific regions exhibiting higher intensity. Intriguingly, the IFCs precisely intersect these regions of higher intensity. The insets offer a closer look at the FFT of the simulated field profile in the higher ${k_x}$ and ${k_y}$ range, revealing shapes that match the analytical IFC and providing crucial insights into the material's support for high $\textit{k}$ vectors. This comprehensive analysis underscores the presence of hyperbolic nature of the polaritons within the structured $\alpha$-MoO$\textsubscript{3}$. \par

\subsection*{Tuning of HPhPs using different $\alpha$-MoO$\textsubscript{3}$ nano structures}

\begin{figure*}[htp!]
    \includegraphics[width=\linewidth]{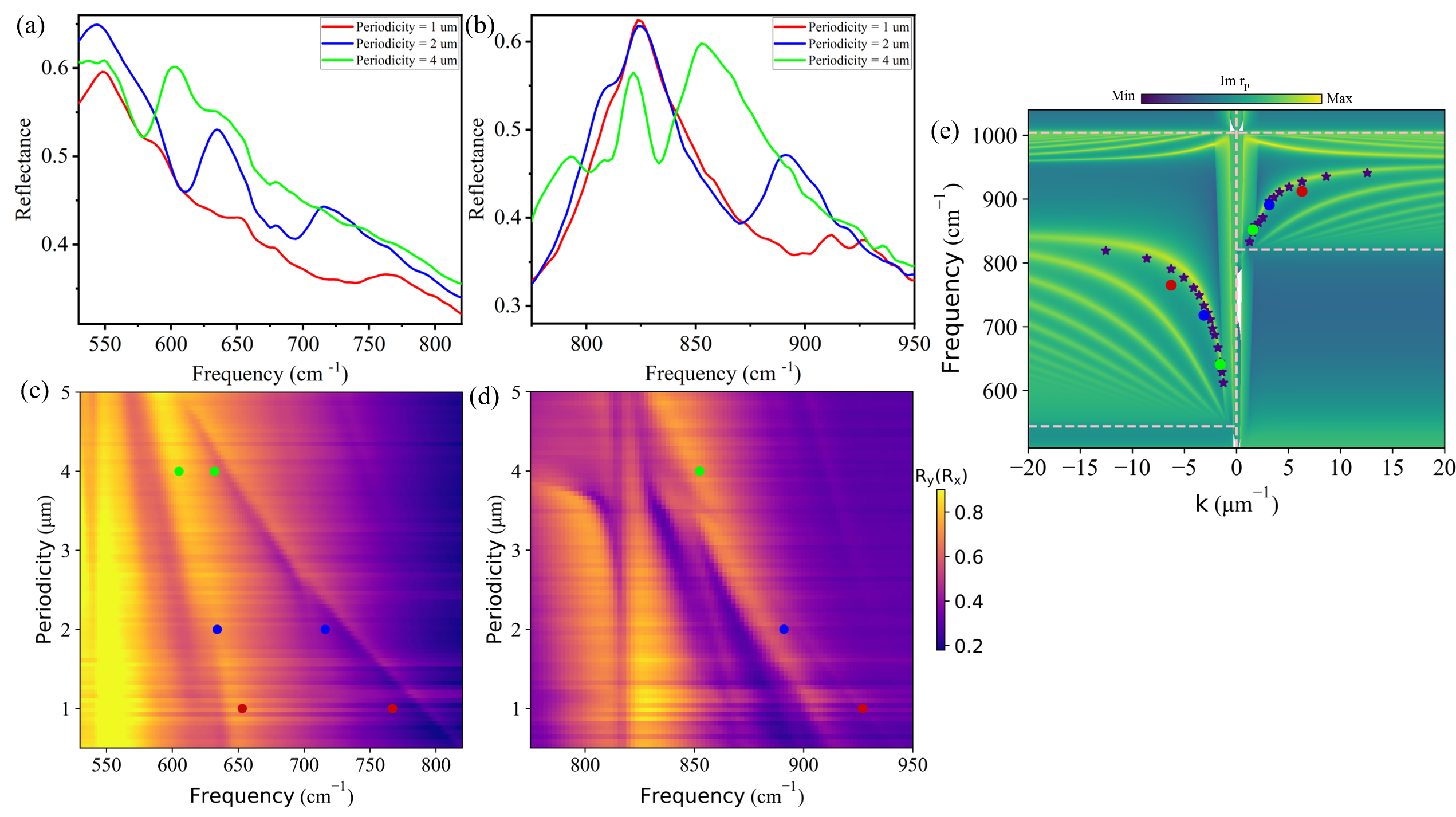}
    \caption{\small \textbf{{Tuning of PhP due to $\alpha$-MoO$\textsubscript{3}$ hypercrystal}}: The experimental results illustrate the far-field reflectance spectra of three cylindrical nanohole structures with periodicity of $\SI{0.5}{\micro\metre}, \SI{1}{\micro\metre}$ and $\SI{2}{\micro\metre}$, respectively, in (a) RS band 1 (for s-polarized incident light) and (b) RS band 2 (for p-polarized incident light) for a 590 nm thick $\alpha$-MoO$\textsubscript{3}$ crystal. The plot highlights a significant resonance shift due to changes in the hole periodicity. For a wide range of hole periodicity in (c-d), the simulated far-field reflectance plotted for both polarized incident light in their respective RS bands. The experimental reflectance peak position is shown as colour sphere in the simulated colour plot, showing experimental results almost match the simulation. Additionally, the simulated and experimentally obtained reflectance peak position (first-order) for different periodicity are shown in (e) through a theoretically calculated HPhPs dispersion plot. The false color plot is calculated by plotting the imaginary part of reflectivity for different frequencies and momenta for a 590 nm $\alpha$-MoO$\textsubscript{3}$ film. The simulated and experimental points are represented by star and sphere, respectively, showing excellent agreement with the theoretical hyperbolic polaritonic dispersion. }
\end{figure*}

The previous experimental results have shown that HPhPs can be excited in $\alpha$-MoO$\textsubscript{3}$ in the far field using a hypercrystal. In order to demonstrate the tunability of HPhPs resonance in this material, three different sets of hypercrystal with different hole sizes of $\SI{0.5} {\um}$, $\SI{1} {\um}$, and $\SI{2} {\um}$ with periodicity of $\SI{1} {\um}$, $\SI{2} {\um}$, and $\SI{4} {\um}$, respectively, were fabricated. For s-polarized incident light, along which the y[001]-crystal direction permittivity($\epsilon_{y}$) is negative, the HPhPs in RS band 1 were excited. A clear blue shift in the HPhPs resonance of the reflectance spectrum was observed from 735 cm$\textsuperscript{-1}$ to 632 cm$\textsuperscript{-1}$ as the nano hole periodocity increases from $\SI{1} {\um}$ to $\SI{4} {\um}$, which is shown in Fig. 3(a). Similar behaviour was observed for higher order modes in band 1 as well where the reflectance peak is having a blue shift from 653 cm$\textsuperscript{-1}$ to 605 cm$\textsuperscript{-1}$ with increasing in the periodicity. Similarly, for p-polarized incident light, along which the x[100]-crystal direction permittivity($\epsilon_{x}$) is negative, the HPhPs in RS band 2 were excited. A blue shift in the reflectance resonance peak was observed from 927 cm$\textsuperscript{-1}$ to 852 cm$\textsuperscript{-1}$ while increasing the nanohole periodicity from $\SI{1} {\um}$ to $\SI{4} {\um}$, respectively (Fig. 3(b)). Each resonance peak in RS band 1 and 2 for different nano hole periodicity corresponds to HPhPs, which is evident from its simulated electric field profile ( see Supplementary information). The simulated far-field reflectance spectra for different nano hole periodicity is plotted Fig. 3(c-d) for both RS bands. The coloured spheres represent the experimental reflectance peak position, almost matching the simulation results and following the blue shift. Small discrepancies occur for smaller nano hole diameters due to anomalies in the shape of the experimentally obtained cylindrical nano holes compared to the ideal cylindrical nano hole in the simulation. The nanoholes were slightly truncated on the bottom side and irregularly shaped. Additionally, the effect of the knife-edge aperture of the FTIR affected the spectrum. Further, the resonance shift arises due to the additional momentum given to the HPhPs by varying the hypercrystal. To understand this, we model the additional momentum by these hypercrystals as being nearly equal to the momentum coming due to a periodic grating which is $\frac{2m\pi}{P}$\cite{gao2012excitation}, where $P$ is the periodicity and $m$ is the difraction order. To validate this approximation, the HPhPs dispersion plot for $\alpha$-MoO$\textsubscript{3}$ was plotted, which is theoretically the false-color plot of the imaginary part of the reflectivity (r$\textsubscript{p}$), as shown in Fig. 3(e). The dispersion relation represents the support of high $\textit{k}$ wave vectors in the hyperbolic regime of $\alpha$-MoO$\textsubscript{3}$. The simulated first-order polariton resonance peak frequencies from both the bands were plotted in the dispersion plot (in both y([001]) and x([100]) directions) for different $\textit{k}$ values ($m=1$) of the hypercrystal (stars), showing excellent agreement with the dispersion plot. Further, the coloured sphere represents the experimental data, which agrees well with the polaritonic dispersion plot. This shows that far-field tuning of the polariton resonance is possible in both the RS bands using the same hypercrystal geometry. In addition to the first order mode in the dispersion, several higher-order slab phonons modes\cite{kumar2015tunable} can be seen in the dispersion plot for both RS bands (Fig. 3(e)). The experimental realisation of these modes is beyond the scope of the present work. 

\par

\subsection*{Tuning of HPhPs by twisting $\alpha$-MoO$\textsubscript{3}$ hypercrystal}

\begin{figure}[htp!]
    \includegraphics[width=\linewidth]{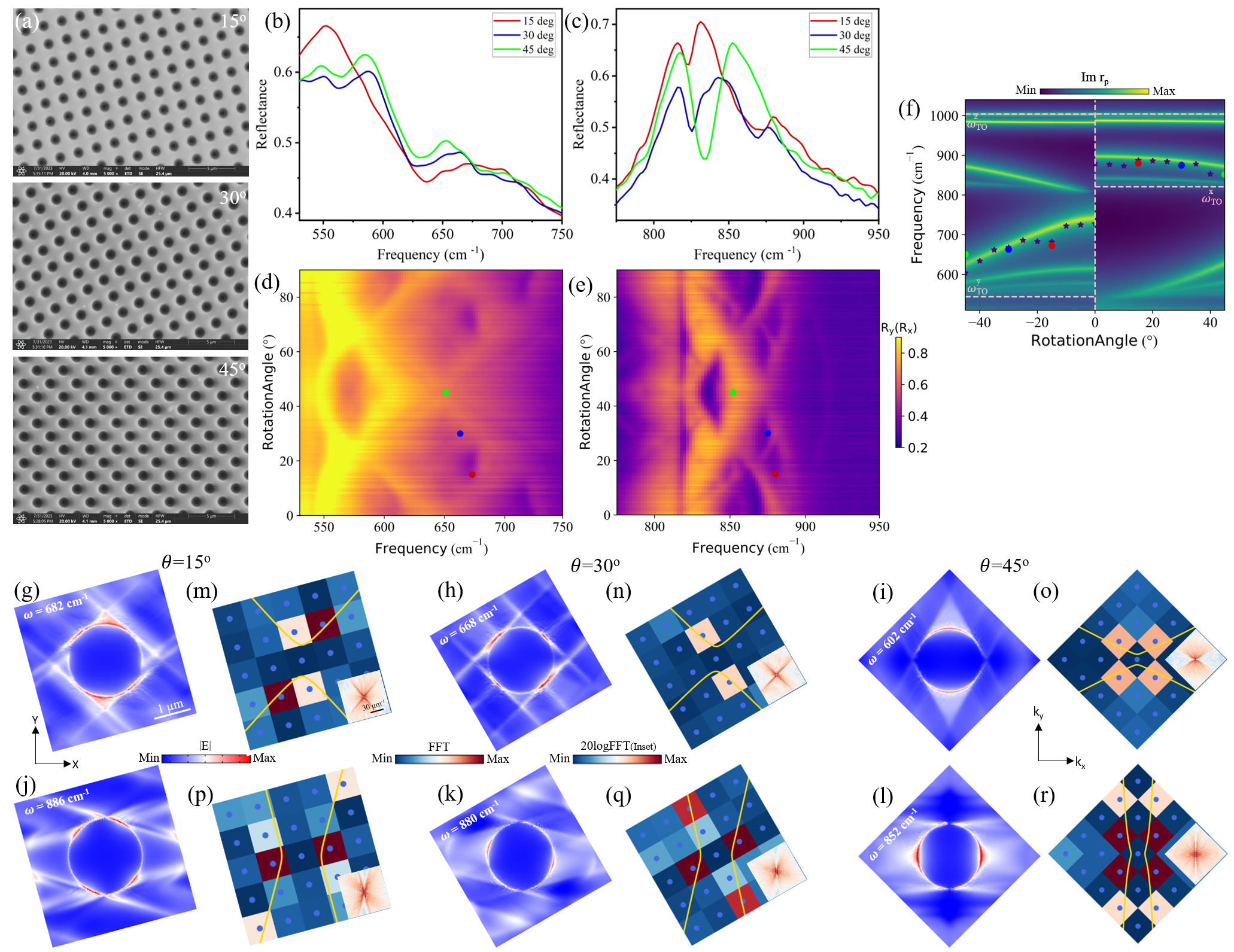}
    \caption{\small \textbf{{Far-field responses of twisted $\alpha$-MoO\textsubscript{3} hypercrystal }}: (a) SEM images depict $\alpha$-MoO\textsubscript{3} hypercrystal with twist angles of 15$^{\circ}$, 30$^{\circ}$, and 45$^{\circ}$, respectively, with periodicity of $\SI{2} {\um}$. Experimental far-field responses for $s$- and $p$-polarized incident light are shown in (b) and (c), illustrating HPhPs tuning with different twist angles within respective RS bands. In (d-e) simulated reflectance color plots with a large number of twist angles are presented in their respective incident polarized light within both RS bands. The experimental resonance peaks are denoted by coloured spheres in the simulated color plot. (f) shows theoretically calculated false color plot by plotting the imaginary part of reflectivity for different frequencies and twist angle for a fixed periodicity of $\SI{2} {\um}$. The simulated and experimental points are represented by star and sphere, respectively, almost matching the theoretical hyperbolic polaritonic dispersion for a wide variety of twist angles. Twist angles above 45$^{\circ}$ are not plotted due to the symmetric nature of the proposed hypercrystal, which is also evident from the simulated colour plot. Inplane ($xy$ plane) electric field profiles ($|E|$) at various resonance frequencies and three different twist angles are shown for $y$-polarized incident light in (g-i) and for $x$-polarized incident light in (j-l), respectively. The $xy$ plane is located 5 nm above the $\alpha$-MoO$\textsubscript{3}$ hypercrystal. (m-r) shows the nature of analytically calculated IFCs for specific resonance frequencies corresponding to Fig(g-l). These hyperbolic IFC are plotted on top of respective FFT images of the corresponding real part of ${E_z}$, showing intersection of IFCs with higher intensity FFT profiles giving different mode orders. The blue dots represent the reciprocal space lattice points by considering a periodic lattice with periodicity $\SI{2}{\um}$. Further, the inset represents the FFT of the simulated field profile for higher $\textit{k}$ value confirming propagation of HPhPs at oblique angles. For the experiment and simulation, a 560 nm $\alpha$-MoO$\textsubscript{3}$ film is considered.}
\end{figure}

Here, we present a comprehensive exploration of the impact of rotation of the $\alpha$-MoO\textsubscript{3} hypercrystal and their corresponding far-field responses. For this, we choose three different $\alpha$-MoO\textsubscript{3} hypercrystal with rotation angles ($\theta$) of 15$^{\circ}$, 30$^{\circ}$ and 45$^{\circ}$, all possessing a periodicity of $\SI{2} {\um}$ with hole diameter $\SI{1} {\um}$. The SEM images in Fig. 4(a) offer a visual representation of the fabricated patterns which shows that the hypercrystal is rotated at different $\theta$ with respect to [100] ($x$-axis) of the $\alpha$-MoO\textsubscript{3} crystal. When the hypercrystal for a fixed periodicity is rotated, the wavevector of HPhPs also rotates. This rotation enables the excitation of HPhPs with different orientations within the basal plane of the $\alpha$-MoO\textsubscript{3} crystal. The resonance frequency of the HPhPs relies on the orientation of their wavevector, which is influenced by the inplane hyperbolic dispersive nature $\alpha$-MoO\textsubscript{3} and consequently, the HPhP resonances in the patterns are significantly influenced by the rotation angle $\theta$. This is in contrast to in-plane isotropic crystals made of graphene or hBN, where these low symmetry modes are symmetric and cannot be tuned by rotation. The experimental far-field responses for three different $\theta$ is shown in Fig. 4(b-c), whereas the simulated far-field reflectance for a wide range of different $\theta$ are presented in Fig. 4(d-e), for both RS bands with $y$- and $x$- polarized incident light respectively. These graphs effectively illustrate the tuning of HPhPs with different $\theta$ values due to the change in the spatial arrangement of the holes. Further, the angle-dependent behaviours can be seen more clearly by plotting the dependency of resonance frequency with $\theta$ for both RS bands in Fig. 4(f), which agrees well with the calculated colorplot of the imaginary part of reflectivity. The simulated (star) and experimental data (sphere) agree well with the analytical hyperbolic dispersion, thus giving us more degrees of freedom for the excitation of HPhPs at various frequencies for a fixed periodicity. For a deeper understanding of the system, we plot the normalized electric field profile at various resonance frequencies for both RS band 1 (Fig. 4(g-i)) and RS band 2 (Fig. 4(j-l)), respectively, showing the nature of inplane polariton propagation against the rotation of hypercrystal. IFC of the corresponding frequencies and $\theta$ are plotted in Fig. 4(m-r). The hyperbola major axis is oriented along ${k_y}$ (RS band 1) and ${k_x}$ (RS band 2), enabling excitation of HPhPs via twisting the hypercrystal. The FFT image of the simulated real part of $E_z$, shows the associated resonance modes where the IFC hit the maximum FFT intensity profile. Further, the inset shows FFT for high $k$ values, which matches with analytical IFC plot, hence describing the rich inplane hyperbolic propagation of polaritons at an oblique angle. Thus, rotating the hypercrystal allows for precise control over the wavevector and resonance frequency of the HPhPs in both RS bands. This exceptional and influential attribute of $\alpha$-MoO\textsubscript{3} hypercrystal can be utilized to create various groundbreaking optical devices.

\section*{DISCUSSION}

In this comprehensive study, we explored the intriguing realm of far-field excitation of phonon polaritons using nanostructured in-plane hyperbolic van der Waals crystals. Through the meticulous construction of photonic hypercrystals, we achieved the successful elicitation of in-plane HPhPs in the far field and demonstrated a large of tunability by deftly manipulating the hypercrystal lattice. This pivotal advancement effectively overcame the significant difference in momentum between regular light and polaritons in open space, thus clearing the way for the practical integration of HPhPs into various innovative devices. Our exhaustive experimental and simulation results conclusively confirm that the additional resonances observed were unequivocally attributed to HPhPs. Moreover, the strategic rotation of $\alpha$-MoO\textsubscript{3} hypercrystals unveiled an extraordinary degree of control over wavevectors and resonance frequencies. The intrinsic properties of $\alpha$-MoO\textsubscript{3} within its surface plane, characterized by in-plane biaxial hyperbolic behavior, position it as a standout candidate for further explorations in the domain of polaritonic waves. The IFC at different polaritonic frequencies provided concrete evidence of the hyperbolic nature of polaritons within the structured $\alpha$-MoO\textsubscript{3}, presenting a wealth of opportunities for innovative device design. This research advances our fundamental understanding of the intricate interplay between light and hyperbolic materials in the far-field range and establishes a robust foundation for practically implementing phonon polariton-based technology. The discoveries made in this study are set to pave the way for a new phase in advancing functional and adaptable optical devices, offering a wealth of possibilities for refining polariton behavior.

\section*{MATERIALS AND METHODS}

\subsection*{Synthesis of $\alpha$-MoO\textsubscript{3} thin film}
The synthesis of $\alpha$-MoO\textsubscript{3} was achieved through the physical vapor deposition method utilizing a single-zone furnace and two quartz tubes of differing sizes. The inner tube, measuring 2 cm in diameter and 20 cm in length, housed 100 mg of 99.9\% pure $\alpha$-MoO\textsubscript{3} (Sigma Aldrich) at the zone center of the furnace, while the larger tube measured 8 cm in diameter and nearly 90 cm in length. The heating zone was maintained at 710 \textsuperscript{o}C with a heating rate of 19.7 \textsuperscript{o}C per minute, gradually increasing to 750 \textsuperscript{o}C over 20 minutes before cooling down to room temperature naturally. Atmospheric pressure was maintained throughout the process, with Argon gas supplied at 60 sccm.\cite{sun2019one} The resultant free-standing $\alpha$-MoO\textsubscript{3} flakes were collected from the inner quartz tube, followed by mechanical exfoliation to transfer the as-grown single crystal $\alpha$-MoO\textsubscript{3} to the Si substrate which involved thermal release tape for exfoliation, and the thermal tape-$\alpha$-MoO\textsubscript{3}-Si substrate system was heated at 100 \textsuperscript{o}C for few minutes to obtained the desired samples. The thickness of the exfoliated flakes measured with a Bruker's DekTak XT stylus profilometer. \par

\subsection*{Fabrication of $\alpha$-MoO\textsubscript{3} hypercrystal}

To fabricate $\alpha$-MoO$\textsubscript{3}$ nanostructures, we used dual beam Focused Ion Beam (FIB) etching system (Helios 5 UC from Thermoscientific) equipped with Scanning Electron Microscopy (SEM)(Thermoscientific) for high-resolution imaging purposes. The 30 kV Ga\textsuperscript{+} FIB with a beam current of 2.5 nA and beam diameter $\approx$ 136 nm used to create nanostructures. We fabricated nano structure with different sizes by creating a bmp file of the desired geometry in a $\SI{200} {\um}$ $\times$ $\SI{200} {\um}$ rectangular area with the required number of pixels. The horizontal pixel resolution was set at 2741 pixels with 2500x magnification during the patterning. After the desired $\alpha$-MoO$\textsubscript{3}$ nanostructures were created, samples were annealed for 2.5 hours at 300 deg \textsuperscript{0}C at O\textsubscript{2} ambient condition to eliminate Ga\textsuperscript{3+} intercalation and obtain a better PhPs response.\cite{zheng2019mid,huang2021van} For the far-field PhP study, we created cylindrical nano holes with different hole diameters (with periodicity twice the hole diameter) and with different Moo3 thicknesses at three rotation angles(15$^{\circ}$, 30$^{\circ}$, 45$^{\circ}$),  which described in the results section.

\subsection*{Far field  measurements of $\alpha$-MoO\textsubscript{3} hypercrystal using micro-FTIR}

The Bruker Hyperion-3000 FTIR microscope, equipped with a Bruker Vertex 80 spectrometer, was used to study the far-field optical response of $\alpha$-MoO\textsubscript{3} nano patterns in the 545 cm\textsuperscript{-1} -- 1050 cm\textsuperscript{-1} spectral region. A 15X Cassegrain objective with a NA of 0.40 and an average off-normal incident angle of about 17$^{\circ}$ was used to illuminate the IR light. The spectrum was recorded using a wideband MCT detector with a spectral resolution of 2 cm\textsuperscript{-1} and 256 number of scans. The flakes were oriented such that the $x$-axis of the FTIR instrument is parallel to the X-axis ([100] direction) of $\alpha$-MoO\textsubscript{3} crystal. The reflectance spectra were collected with respect to the gold background in the back-scattering geometry with a broad mid-IR incident plane polarized light using a KRS-5 based IR polarizer. To select an area on $\alpha$-MoO\textsubscript{3} nano structures, a knife-edge aperture was set with an area of $\SI{200} {\um}$ $\times$ $\SI{200} {\um}$. Finally, to remove noise, the FTIR spectra for p-polarized light were smoothened by weighted adjacent averaging from 9 points.

\subsection*{Analytical calculation and full-wave numerical simulations}

The analytical calculations were done using the transfer matrix method by considering a thin infinite $\alpha$-MoO\textsubscript{3} slab on top of a silicon substrate. The dielectric tensor of $\alpha$-MoO\textsubscript{3} in the mid-IR spectral region can be given using a multiple Lorentz oscillator model\cite{alvarez2020infrared}:

\begin{equation}
    \epsilon_i = \epsilon_{\infty,i}\prod_{j=1}^{N_i} \left(1 + \frac{(\omega_{ij}^{LO})^2 - (\omega_{ij}^{TO})^2}{(\omega_{ij}^{TO})^2 - \omega^2 - \imath\omega\Gamma_{ij}}\right)
\end{equation}

where, $N_i$ is the number of oscillators contributing in the $i$th direction, $\varepsilon_{i}$ is principle component of dielectric tensor and $i = x, y$ and $z$ represent [100], [001] and [010] crystallographic axes in the $\alpha$-MoO\textsubscript{3} crystal. $\varepsilon_{\infty,i}$ is high frequency dielectric constant, $\omega_{ij}^{LO}$ and $\omega_{ij}^{TO}$ represent the frequency of longitudinal optical (LO) and transverse optical (TO) phonons along $i{\text{th}}$ direction respectively and $\Gamma_{ij}$ represents the line-width. Further, the IFCs were plotted by using a quasi-static approximation, as the dispersion relation of polaritons in a biaxial slab can be given using the following equation,\cite{alvarez2019analytical} 

\begin{equation}
    k(\omega) = \frac{\psi}{t}[arctan(\frac{\psi\epsilon_{1}}{\epsilon_{z}}) + arctan(\frac{\psi\epsilon_{3}}{\psi\epsilon_{z}}) + \pi l]
\end{equation}
 where the phonon mode order $l$ = 0,1,2.., the inplane wave vector $k =\sqrt{k_x^2+k_y^2} $  , t is the thickness of the biaxial slab, $\epsilon_
 {1},\epsilon_{3}$ is the permittivity of air and silicon respectivily.  $\psi= \iota \sqrt{\frac{\epsilon_{z}}{\epsilon_{x}cos^2(\phi)+\epsilon_{x}sin^2(\phi)}}$ , where $\phi$ is the angle between $k$ and $x$ axis.  Here, in this calculation we choose the nearmost hyperbola ($l$=0) and $|Re(k)|>Im(k)$. 

 To accurately model the far-field polaritonic resonance observed in our experimental studies, we utilized full-wave numerical simulations employing the finite boundary element method in COMSOL Multiphysics. The simulations used to calculate the far-field reflectance, $|E|$  and the near-field Re($E_{z}(x, y))$. Our modelling focused on a unit cell consisting of a cylindrical nano hole of $\alpha$-MoO\textsubscript{3} on top of Si substrates, with a variable hole diameter and, periodicity twice of hole diameter, with thickness of 590nm (for periodicity variation) and 560 nm (for twisting). Both the incident and collection ports were positioned well above the $\alpha$-MoO\textsubscript{3} nano structures for far-field studies. The incident wave was modelled as a plane-polarized periodic electric field which was incident normally. Our simulations were essential in accurately modeling the observed polaritonic resonances in the far-field, and allowed us to better understand the optical response of our system.

\bibliography{scibib}

\bibliographystyle{Science}

\section*{Acknowledgments}
A.K. acknowledges funding support from the Department of Science and Technology via grant number CRG/2022/001170. N.R.S. acknowledges the Council of Scientific $\&$ Industrial Research fellowship No: 09/087(0997)/2019-EMR-I. P.S. and B.K. acknowledges support from Prime Minister's research fellowship (PMRF), Government of India.

\clearpage

\end{document}